# Modeling of Electrochemical Oxide Film Growth – A PDM Refinement

Ingmar Bösing*[1,3], Fabio La Mantia[2,3], Jorg Thöming[1,3]

[1]Chemical Process Engineering Group (CVT)
[2]Energy storage and conversion systems
[3]University of Bremen
*corresponding author, ingmarboesing@uni-bremen.de
Leobener Strasse 6
28359 Bremen
Germany

*Abstract* The Point Defect Model (PDM) is known for over 40 years and has brought deeper insight to the understanding of passivity. During the last decades it has seen several changes and refinements, and it has been widely used to analyze growth kinetics of different alloys. Nevertheless, the model has been based on still unconfirmed assumptions, as constant and potential independent electric field strength. To overcome this limitation, we introduce a Refined PDM (R-PDM) in which we replace those assumptions by using additional equations for charge distribution including new physically valid boundary conditions based on considering finite dimensions for the defects by introduction of two defect layer at the film boundaries and by calculating the potential drop at the surface of the film towards the solution over the compact double layer. The calculations by the R-PDM show that the original PDM assumptions are only valid for very specific parameter combinations of oxide film growth and vacancies transport and cannot generally be taken for granted. We believe our findings of electric field and potential drop dependency on the external potential to pave the way for a more realistic description of passive layer formation.

## 1. Introduction

We are surrounded by metal oxides in our daily live in form of semiconductors [1], corrosion products, and passive films for corrosion protection [2]. The modeling of oxide film growth and properties can deliver a deeper insight into the understanding of electrochemical metal oxide formation. In recent literature and research the Point Defect Model (PDM), developed by Macdonald [3,4], and its extensions [5–7] have been mostly widely explored and used to describe the film growth of oxides. The model is currently widely used for the description of passive film kinetics of different metals and alloys [8–10], the analysis of Mott-Schottky data [11,12] the investigation of corrosion processes [13–17], to describe pitting corrosion [18] and the comparison of passive film protection between different alloys [19], but there are some critical points limiting its applicability [7,20,21].

The PDM describes film growth of oxides as a result of interfacial reactions and defect transport (which gives the name of the model) within the oxide film based on the Nernst-Planck equation. Even though the PDM is widely used it is based on a few assumptions, which have not been fully proved. Among them, two fundamental ones are:

1.) The field strength ($F_E$) inside the oxide film was assumed to be constant and independent of the external potential $\varphi_{ext}$. This was justified as follows: "The electric field strength ($\varepsilon$) (*sic*) is independent of voltage and distance through the film. The former is due to band-to-band (Esaki) tunneling, such that as the field strength tends toward a higher value the bands become steeper and the tunneling distance for electrons from the valence band to the conduction band (or to inter band gap states) decreases, thereby resulting in a higher tunneling current. This produces a separation of charge that opposes the field and essentially buffers the field at an upper, voltage- independent value of 2–5 MV/cm" [4].

2.) The potential drop at the oxide/solution interface was assumed to be a linear function of the external potential and pH: $\varphi_{fs} = \alpha \varphi_{ext} + \beta \text{pH} + \varphi_{fs}^0$, where $\alpha$, $\beta$ and $\varphi_{fs}^0$ depend on the type of metal and the environment.

There has been an ongoing debate to which extent these assumptions can be justified and limit the applicability of the PDM. Different approaches have been suggested to overcome such a limitation. In 1995 Battaglia and Newman published a comprehensive model to describe the growth of iron oxide on an iron electrode [22]. They assumed a high field mechanism for the transport of defects in the oxide film and calculated the field strength by the Poisson equation. In 2006 Vankeerberghen published a research paper in which he presented a model to calculate the steady state properties of metal oxides based on the PDM but without the above-mentioned PDM assumptions [20]. While Vankeerberghen assumed the same interfacial reactions and defect transport as seen in the PDM, he used the Poisson equation to calculate the electric field strength inside the oxide film to overcome the PDM assumptions. In his paper he concluded, in contrast to the PDM, that the electric field strength inside the passive film is not constant. He further concluded the potential drop at the film/solution interface, $\varphi_{fs}$, depends linearly on the polarization potential, which is in agreement with the PDM assumption. In his model Vankeerberghen did not incorporate the presence of electronic charge carriers (electrons and holes), even though he discussed a possible case in the appendix. In his thought experiment he underestimated the charge of the electrons and holes by including the elementary charge to the Poisson equation. Therefore, the inclusion of these charge carriers needs to be discussed more in detail. Furthermore, in his model (and models depending on his approach [23]) Vankeerberghen assumed that the electric field strength to be 0 at the oxide film boundaries ($F_E(x=0) = 0$ and $F_E(x=L) = 0$), which is physically not plausible. In 2012 Albu et al. published a research paper based on the Vankeerberghen model which investigates the contribution of the solution in front of the oxide film to the film growth and properties [24]. Even though they did not discuss the boundary conditions of the Poisson equation, they stated that the model is based on the Vankeerberghen model. This leads to the assumptions that they used similar boundary conditions. Furthermore, Albu et al. ignored the diffusion double layer region in the electrolyte, which could, on the opposite, affect strongly the kinetic of film growth.

In 2010 Bataillon et al. also presented a model for the calculation of oxide films based on interfacial reactions and defect transport, which includes the Poisson equation for the calculation of the electric field [25]. In contrast to the presented model, Bataillon et al. assumed the reactions at the interfaces to be reversible (which is a further difference to the classic PDM) and they used Gauss law for the description of the boundary conditions for the Poisson equation. The, so-called Diffusion Poisson Coupled Model (DPCM) assumes a surface charge at the film interfaces, which leads to an electric field inside the film. The electric field can be calculated by the differential capacitance of the interfaces and their potential of zero charge, which need to be guessed for the calculation. We have preferred a different approach in which, as we will show, we are modeling the differential capacitance of the interface in a similar fashion as it is done in the double layer theory [26].

A generalized growth model was presented by Seyeux et al. in 2013 [7]. They considered additional possible reaction equations and a time depending electric field strength. In contrast to our model they assumed the electric field to be constant over the whole length of the oxide film and the potential drop at the metal/film interface to be constant during growth.

In 2019 Engelhardt et al. calculated the electric field strength inside the oxide films based on the transport equations and the Poisson equation and found it to be constant over a broad range of the film for sufficient thin films without the incorporation of electrons and holes [27]. For thick oxide films they found the electric field to be not constant. Their calculation was based, amongst other things, on the assumption of the linear dependency of the film/solution potential drop $\varphi_{fs}$ on the external potential and the pH.

In this work we demonstrate that, by extending the number of equations of the PDM without increasing the number of parameters, we can eliminate the two discussed assumptions. These extensions allow for calculating the potential distribution over the oxide film and the potential drop at the film solution interface by using the Poisson equation as well as the transport of defects by the Nernst-Planck equation.

In contrast to previous work, the boundary conditions introduced in our model take into account the physically dimensions of ions and defects and present an explanation for the potential drop at the film interfaces. By means of the presented comprehensive Refined PDM (R-PDM) we calculate the film thickness and properties of an oxide film growing on a metal. The model is based on the interfacial reactions given by the PDM (whereas only 5 of the 7 given reactions are used for simplification. The additional reactions can be easily added). The influence of the polarization potential on the average defects and electric field, as well as the defect and electric field distribution will be analyzed and discussed in details. Furthermore, a comparison to calculations of film thickness and defect concentrations based on the classic PDM is given.

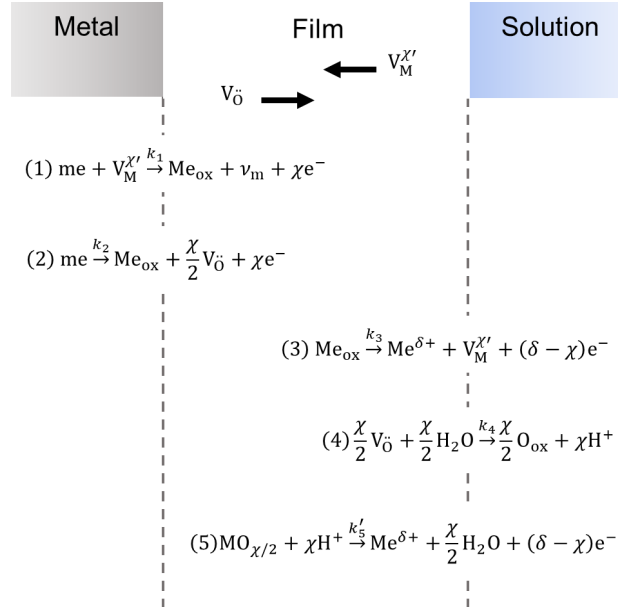

Figure 1: Reaction system. me: metal ions in the metal, $V_M^{\chi'}$: metal vacancies with positive charge $\chi$, $v_m$: metal vacancy in the metal, e: electron, me: metal atoms in the metal, Me$_{ox}$: metal ion in the oxide, $V_{\ddot{O}}$: oxygen vacancy two times positively charged, $V_M^{\chi'}$: metal vacancies $\chi$ times negatively charged, Me$^{\delta+}$: Metal ion in solution, O$_{ox}$: oxygen ion in the oxide.

## 2. MODEL DESCRIPTION

### 2.1 Interfacial reactions

The PDM introduces 7 interfacial reactions which lead to film growth, film dissolution and the production and consumption of oxygen vacancies, metal vacancies and metal interstitials. For the purpose of simplification, we reduce the model by two interfacial reactions (namely the production and consumption of metal interstitials, see Figure 1). These two additional reactions can, if necessary, in more complex cases, also implemented in the model.

Two reactions occur at the metal/film (mf) interface: Reaction 1, the consumption of metal vacancies $V_M^{\chi'}$ (CMV), and Reaction 2, the production of oxygen vacancies $V_{\ddot{O}}$ (POV). Reaction 2 is a so-called non-lattice conservative reaction, because the production of oxygen vacancies and metal ions in the oxide leads to the growth of the oxide film into the metal.

At the film/solution (fs) interface occur three different reactions. Reaction 3, the production of metal vacancies (PMV), Reaction 4 the consumption of oxygen vacancies (COV) and the film dissolution reaction (FDR), Reaction 5. The FDR is also a non-lattice conservative reaction, because due to the film dissolution the fs boundary moves and the oxide film shrinks.

If both non-lattice conservative reactions are in equilibrium - moving the interfaces equally - the system reaches steady state. The metal is still dissolving at the mf interface and the film is still dissolving at the fs interface in steady state but the growth of the oxide film (by metal dissolution and oxygen vacancies production) equals the dissolution of the film at the fs interface.

## 2.2 Reaction kinetics

The kinetics of the different reactions can be described by the rate constants:

$$k_i = k_i^{0\prime} \exp\left(\frac{\alpha_i \chi F}{RT}(\varphi_{\text{mf}} - \varphi_{\text{eq}})\right) = k_i^0 \exp\left(\frac{\alpha_i \chi F}{RT}\varphi_{\text{mf}}\right), i = 1,2 \qquad (1)$$

For the two reactions at the mf interface. $k_i^0$ is the standard reaction rate constant of the Reaction $i$ depending on the activation energy of Reaction $i$, $\alpha_i$ is the charge transfer coefficient of Reaction $i$, $n$ is the number of electrons involved in the reaction, $F$ the Faraday constant (96485 As/mol), $R$ the gas constant (8.314 J/mol/K) and $T$ the temperature (293 K). The potential drop at the mf interface $\varphi_{\text{mf}}$ drives the reaction and depends on the film thickness, the defect distribution over the film and the external potential inside the metal. Because it has been pointed out as a critic on this kinetic assumption, that it would not depend on potentials such as the equilibrium potential $\varphi_{\text{eq}}$ or the corrosion potential [28] we would like to point out, that the equilibrium potential is a constant and can be included to $k_i^0$ easily as: $k_i^{0\prime} \exp\left(\frac{\alpha_i nF}{RT}(\varphi_{\text{mf}} - \varphi_{\text{eq}})\right) = k_i^0 \exp\left(\frac{\alpha_i nF}{RT}\varphi_{\text{mf}}\right)$. A detailed description of the potential distribution in the system is given below. Furthermore, as in the original PDM we assume the interfacial reactions to be irreversible. This simplification can be justified in this simulated case by the assumption of the reaction being far from equilibrium. Nevertheless, treating the reactions as reversible is also a possible case, as shown by Bataillon [25]. The question, which case fits more to the realistic behavior of investigated alloys, still needs to be evaluated.

The rate constants of Reaction 3 and Reaction 4 is described by:

$$k_i = k_i^0 \exp\left(\frac{\alpha_i(\delta - \chi)F}{RT}\varphi_{\text{fs}}\right), i = 3,4 \qquad (2)$$

$\varphi_{\text{fs}}$ describes the potential drop at the film/solution interface. The dissolution of the oxide film (Reaction 5) is assumed to be a chemical – potential independent – reaction which depends on the concentration of protons $c_{\text{H}^+}$ at the fs interface and the kinetic order of film dissolution $n$.

$$k_5 = k_5^0 c_{\text{H}^+}^n \qquad (3)$$

According to the described kinetic constants the steady state condition for oxide film growth can be described by:

$$\frac{dL}{dt} = 0 = \Omega(k_2 - k_5) \tag{4}$$

This steady state condition with the molar volume of the oxide $\Omega$ can be used to determine the steady state film thickness as described in Section 3.1.1.

## 2.3 Transport equations inside the oxide

The transport of the defects can be described by the Nernst-Planck equation for the one-dimensional case:

$$J_i = -D_i \frac{\partial c_i}{\partial x} - \frac{z_i F D_i}{RT} \frac{\partial \varphi}{\partial x} c_i \tag{5}$$

With the flux of moles $J_i$ of species $i$, namely the oxygen vacancies and metal vacancies, the concentration $c_i$, the diffusion coefficient $D_i$ of species $i$, the charge $z_i$ of species $i$ and the potential $\varphi$.

### 2.3.1 Boundary conditions

The transport of metal vacancies and oxygen vacancies and the boundary conditions at the interfaces is schematically represented in Figure 2. The boundary conditions for the flux of metal vacancies $J_{MV}(x = 0)$ and oxygen vacancies $J_{OV}(x = 0)$ at the metal/film interface are:

$$J_{MV}(x = 0) = -k_1 c_{MV} \tag{6}$$
$$J_{OV}(x = 0) = k_2 \tag{7}$$

with $c_{MV}$ the concentration of metal vacancies. The boundary conditions at the film/solution interface are:

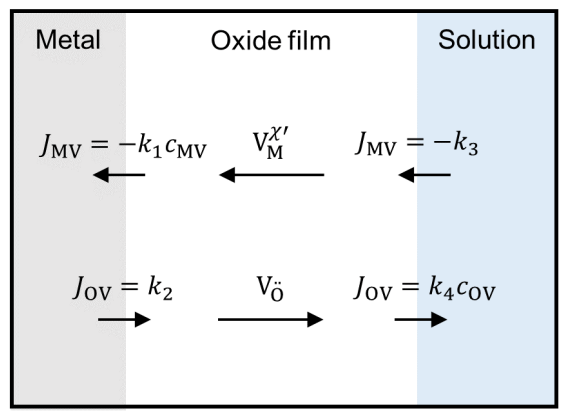

Figure 2: Schematic view of the boundary condition for the transport of defects inside the oxide film.

$$J_{MV}(x=L) = -k_3 \tag{8}$$

$$J_{OV}(x=L) = k_4 c_{OV} \tag{9}$$

with $c_{OV}$ the concentration of oxygen vacancies.

## 2.4 Potential distribution

The distribution of the potential $\varphi$ over the oxide and the solution can be described by the Poisson equation for the one-dimensional case:

$$\frac{\partial^2 \varphi}{\partial x^2} = -\frac{\partial F_E}{\partial x} = -\frac{F}{\varepsilon_r \varepsilon_0} \Sigma z_i c_i \tag{4}$$

The permittivity of the vacuum $\varepsilon_0$ is 8.85×10$^{-12}$ As/V/m, the relative permittivity $\varepsilon_r$ is material depending, $z_i$ is the charge of the charged species, metal and oxygen vacancies (and electrons and holes in the extended model), and $c_i$ is the concentration of the charged species at the position x.

### 2.4.1 Interface and Boundary conditions

The potential distribution across the metal, the oxide and the solution is schematically shown in Figure 3a, and a detailed view of the defect layer is given in Figure 3b. The defect layer takes into account the dimension of ions and defects. It represents the accumulation of defects at the metal/film interface, which is located between the oxide film, and the metal. It is a charge free space with the thickness $d_{dl}$ which is equal to the lattice constant. Due to the charge free space between the metal and the film the electric field is constant inside the defect layer, as can be seen by the Poisson equation. The potential drop at the metal/film interface, which drives the electrochemical reaction at this interface, is the potential drop over the defect layer. According to the electric flux boundary condition [29] the electric field strength inside the defect layer $F_{E,1}$ and the field strength at the film side of the defect layer/film interface $F_{E,2}$ are connected by:

$$\varepsilon_1 F_{E,1} = \varepsilon_2 F_{E,2} \tag{10}$$

Where $\varepsilon$ is the permittivity, $\varepsilon = \varepsilon_0 \varepsilon_r$ and the indices 1 and 2 represent the left-hand side and right–hand side of the interface, respectively. The potential drop at the metal/film interface is consequently the difference of the external potential and the potential at this interface $\varphi_{mf} = \varphi_{ext} - \varphi(x=0)$ (note that $x=0$ is the defect layer/film interface).

The mathematical formulation of the boundary condition used in the model is in case of the metal/film side:

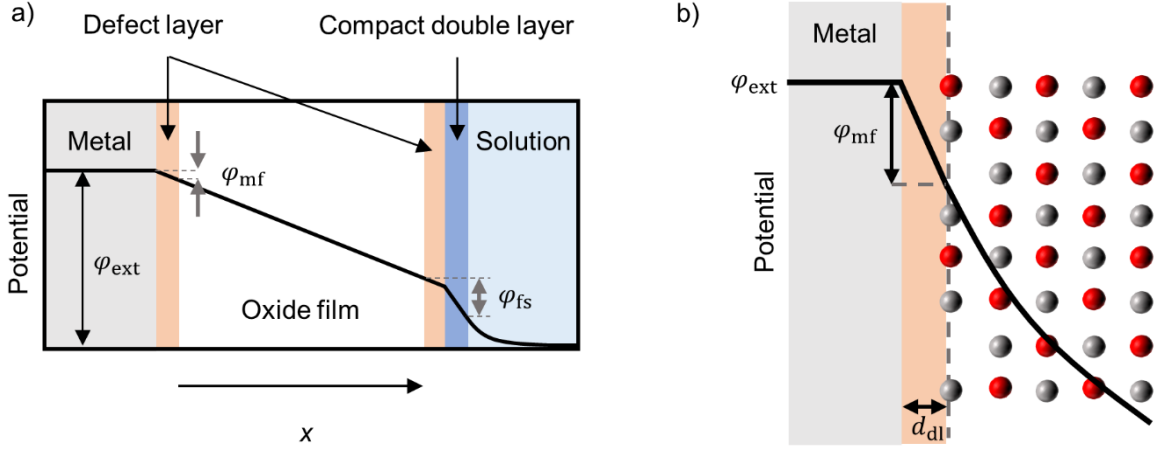

Figure 3: Schematic view on the potential distribution; a) Potential distribution across the metal, film and solution, b) Detailed view of the metal/film interface. Circles represent Ions (red: oxygen, grey: metal). The potential (thick black line) is constant inside the metal and shows a constant slope inside the defect layer which is determined at the defect layer/oxide film interface by the electric flux condition, while the course of potential inside the oxide is defined by the charge densities (and calculated by the Poisson equation).

$$\varepsilon_F \frac{\partial \varphi_F^{mf}}{\partial x} = \varepsilon_{dl} \frac{(\varphi_F^{mf} - \varphi_{ext})}{d_{dl}}, \quad (11)$$

whereas $\varepsilon_F$ and $\varepsilon_{dl}$ represent the permittivity of the oxide film and the defect layer, respectively. $\varphi_F^{mf}$ is the potential inside the film at the metal/film interface and its (negative) partial derivation describes the electric field strength.

At the film/solution interface also a defect layer is formed on the side of the oxide film. Subsequently to the defect layer follows the compact double layer at the solution side. The compact double layer is followed by the diffuse double layer in solution. At these interfaces (film/defect layer, defect layer/compact layer, compact layer/solution) again the continuity of the electric flux boundary condition Equation 10 applies. Inside the defect layer and the compact double layer there is no charge, thus the electric field is constant.

Since the electric field inside the defect layer and the compact double layer is constant, the formulation of the boundary condition at the film solution interface can be written as:

$$\varepsilon_F \frac{\partial \varphi_F^{fs}}{\partial x} = \varepsilon_{dl} F_E^{dl} = \varepsilon_{cdl} F_E^{cdl} = \varepsilon_{sol} \frac{\partial \varphi_{sol}^{fs}}{\partial x}, \quad (12)$$

With the permittivity of the compact double layer $\varepsilon_{cdl}$ and of the solution $\varepsilon_{sol}$. The potential of the solution at the film/solution interface $\varphi_{sol}$ can be calculated by:

$$\varphi_{sol}^{fs} = \varphi_F^{fs} + \frac{\varepsilon_F}{\varepsilon_{dl}} \frac{\partial \varphi_F^{fs}}{\partial x} d_{dl} + \frac{\varepsilon_F}{\varepsilon_{cdl}} \frac{\partial \varphi_F^{fs}}{\partial x} d_{cdl}, \quad (13)$$

whereas $d_{cdl}$ represents the thickness of the compact double layer. For further simplification – assuming a very high ions concentration in the electrolyte - $\varphi_{sol}^{fs}$ can be assumed to be zero meaning that the entire potential drops across the compact double layer (this has not been done in the case of this publication).

One should further note that the reactions at the metal/film interface indeed occur at the defect layer/film interface. For the purpose of this paper the term metal/film interface is interchangeable with defect layer/film interface.

Inside the solution the Poisson equation applies to describe the potential distribution. The potential at $x = L_{\text{cell}}$, i.e. at the end of the cell, is set to 0 to represent the usage of a counter electrode.

## 2.5  Transport equation inside the solution

In our model the transport of cations and anions inside the solution is described by the Nernst-Planck equation. At the compact layer/solution interface a no-flux boundary condition is chosen (the ejection of metal ions to the solution is neglected). This represents the fact that there is no reaction of the ions in solution (the applied potentials are far from the equilibrium potentials of both species). The ejection and transport of the metal ions in the solution caused by the dissolution of the oxide films has been calculated for comparison and did not show any significant effect under the investigated parameters. For simplification the transport and ejection of this species has been neglected. At $x = L_{\text{cell}}$ the ion concentration is fixed to $c_c^0$ for the cations and $c_a^0$ in case of the anions, representing the bulk solution.

## 2.6  Case study parameters

The case study chosen is a pure metal subjected to an acidic liquid environment. We are interested here in predicting the properties of the oxide film being formed on the surface of the metal as well as both potential distribution over the oxide film and transport of vacancies under steady state conditions. For that purpose, we started with educated guesses for the parameters of the R-PDM, which are later varied in case of expected high sensitivity on the outcome or kept constant for all calculations as given in Table 1.

# 3. Results and Discussion

## 3.1  Influence of External Potential

As first variable input parameter the external potential $\varphi_{\text{ext}}$ is varied and the resulting film properties are presented. For this purpose, at first the film thickness for every chosen external potential needs to be determined.

Table 1: Model parameters

| Symbol | Model parameter | Value | Unit |
|---|---|---|---|
| $F$ | Faraday constant | 96485 | C/mol |
| $R$ | Gas constant | 8.314 | J/mol/K |
| $T$ | Temperature | 293 | K |
| $D_{OV}$ | Diffusion coefficient oxygen vacancies | $1 \times 10^{-20}$ | m²/s |
| $D_{MV}$ | Diffusion coefficient metal vacancies | $1 \times 10^{-20}$ | m²/s |
| $D_{cat}$ | Diffusion coefficient cations solution | $1 \times 10^{-9}$ | m²/s |
| $D_{an}$ | Diffusion coefficient anions solution | $1 \times 10^{-9}$ | m²/s |
| $\varepsilon_0$ | Permittivity vacuum | $8.85 \times 10^{-12}$ | F/m |
| $\varepsilon_{cdl}$ | Permittivity compact double layer | 78.5 | |
| $\varepsilon_F$ | Permittivity oxide film | 10 | |
| $z_c$ | Charge number cations | 1 | |
| $z_a$ | Charge number anions | 1 | |
| $z_{OV}$ | Charge number oxygen vacancies | 2 | |
| $z_{MV}$ | Charge number metal vacancies | 2 | |
| $c_c^0$ | Bulk concentration cations | 30 / variable | mol/m³ |
| $d_{cdl}$ | Thickness compact double layer | $10^{-10}$ | m |
| $d_{dl}$ | Thickness defect layer | $5 \times 10^{-11}$ | m |
| $\alpha_i$ | Charge transfer coefficient of Reaction $i$ | 0.1 | |
| $k_1^0$ | Base rate constant of Reaction 1 (CMV) | $5 \times 10^{-8}$ | m/s |
| $k_2^0$ | Base rate constant of Reaction 2 (POV) | $7 \times 10^{-8}$/ variable | mol/m²/s |
| $k_3^0$ | Base rate constant of Reaction 3 (PMV) | $5 \times 10^{-8}$ | mol/m²/s |
| $k_4^0$ | Base rate constant of Reaction 4 (COV) | $5 \times 10^{-8}$ | m/s |
| $k_5^0$ | Base rate constant film dissolution | $7.5 \times 10^{-8}$ | m/s |
| $n$ | Kinetic order film dissolution | 1 | |
| $c_{H^+}$ | Concentration H+ ions | 1 | mol/m³ |
| $\varphi_{ext}$ | External potential | variable | V |

### 3.1.1 Film Thickness

The thickness of the passive film is not a priori known but can be determined by the steady state condition $k_5 = k_2^{ss}$ [20]. According to $k_5 = k_5^0 c_{H^+}^n$ the rate of film dissolution is $7.5 \times 10^{-8}$ m/s. The steady state rate of film growth can be calculated for different oxide film thicknesses. The correct oxide film thickness can be found at the cross section of the steady state rate of film growth and film dissolution (Fig. 4a).

The steady state film thickness shows a linear dependency on the external potential (Figure 4b), which is in agreement with the theory [3] and different measurements [30,31]. The shown calculation of film thickness is in agreement with thickness of different metal oxides, around 2-10 nm. But it is worth to note that the film thickness strongly depends on the choice of different parameters. Alone the choice of a lower film dissolution rate will increase the film thickness significantly as can be seen from Figure 4a (lowering $k_5$ will result in higher $L$ at the cross section). But also, other parameters, as the base rate constant for Reaction 2 (POV) or the ion concentration in solution, affects the film thickness. The values in this manuscript are chosen to achieve typical values of film thickness, electric field and vacancies concentration observed experimentally in oxides formed on stainless steel.

### 3.1.2 Potential and Electric Field

The potential distribution and the electric field inside the oxide film result from the charged defect densities and their distribution, whereas both can be calculated by the Poisson equation. The electric field in oxide films is known to be in the order of magnitude of $10^8$ V/m [11], which is in good agreement with the calculations (Figure 5).

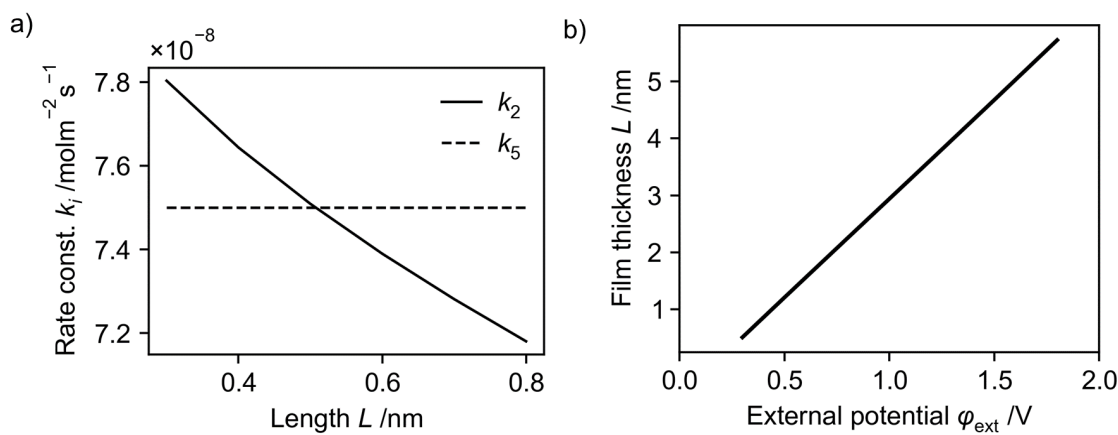

Figure 4: Steady state film thickness $L$ of the calculated oxide film; a) Reaction rate of anion vacancies production ($k2$) and film dissolution (k5) depending on the film thickness for an external potential of 0.3 V. Intersection of both rate constants marks steady state film thickness. b) Steady state film thickness depending on the external potential.

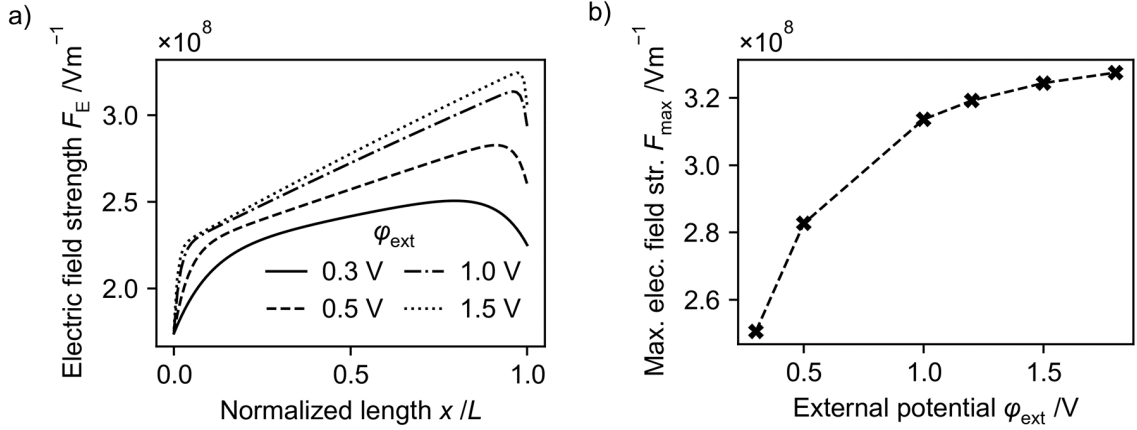

Figure 5: Electric field strength inside the oxide film depending on the external potential; a) Field strength over the normalized film length. b) Maximum field strength inside the passive film.

The electric field is not constant over the oxide film, but shows a steep course at the edges of the oxide (Figure 5a). This is due to the uneven large change in defect concentrations at the oxide film boundaries, which result from the production and consumption of defects at the boundaries. But in contrast to assumptions made by the PDM [4] and some other publications [27,32], the electric field also inside the oxide is not constant. This is due to the unequal charge distribution resulting from the n-doping across the film. If the defect concentration of metal vacancies is equal to the concentration of oxygen vacancies (assuming the same charge number) inside the oxide film, the electric field is constant over a broad range of the oxide film (which can be seen from the Poisson equation). In the model a constant electric field can be achieved by balancing the standard reaction rates accordingly, but it should be noted that in real experiments this is a very unlikely case.

Furthermore, it can be seen that the maximum electric field strength is depending on the external potential (Figure 5b), in contrast with the classic PDM assumption 1), which however is based on the possible field buffering by band-to-band tunnel current due to band-bending by the electric field. The influence of electrons and holes and the interband tunnel current on the electric field will be shown in future work. It just needs to be mentioned that interband tunnel current in thin oxide film ($L < 10$ nm) needs either very high electric fields or very narrow band gaps to show any effect. Considering these observations and the results shown in Figure 5 one can conclude that the electric field is a function of applied potential and of space inside the oxide film. The extend of the dependency of $F_E$ on external potential and space varies with choice of the system parameters.

With increasing external potential, the course of potential shows a much steeper decrease over the passive film (Figure 6a). With increasing external potential, the film thickens and the electric field strength increases. Both lead to a higher potential drop across the film. Due to higher potential drops across the film with increasing external potential the potential drops at the film/solution interface $\varphi_{fs}$ are indeed affected by the external potential (Figure 6b).

### 3.1.3 Vacancies Concentration

The vacancies concentration is dictated by the rate of production and consumption at the boundaries and the electric field. Due to the higher production rate of anion (oxygen) vacancies (which is a function of both the standard rate constant and the potential drop $\varphi_{mf}$) compared to the production rate of cation (metal) vacancies, the concentration of anion oxygen vacancies is higher over a broad range of the oxide (Figure 7a). At the boundary at which the vacancies are consumed by the reactions the concentration decreases sharply. This effect is even more pronounced with higher external potentials (Figure 7b), due to a faster transport of vacancies by migration. Thus, a higher anodic potential leads to a more even vacancies' distribution and a steeper decrease of defect concentration at the interface at which the defect consumption occurs.

The total number of oxygen and metal vacancies lies around $10^{20}$ /cm³ (Figure 8), which is in good agreement with the literature [11]. The amount of Oxygen vacancies is larger compared to the amount of metal vacancies due to the chosen parameters as base rate of oxygen production $k_2^0$.

## 3.2 Influence of Oxygen Vacancies Production Reaction Rate

The base rate constant of the oxygen vacancies production $k_2^0$ does not only affect the concentration of oxygen vacancies but the film growth as well because it dictates the rate of one of the two non-lattice conservative reactions. With increasing base rate of oxygen production, and thus film growth, the steady state film thickness increases (Figure 9).

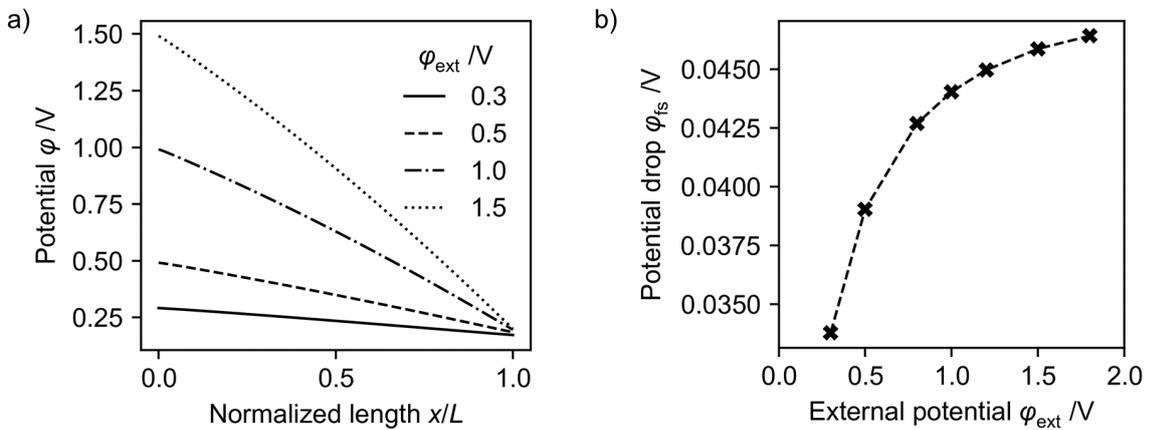

Figure 6: Potential distribution; a) Potential through the oxide film over the normalized film thickness

for different external potentials. b) Potential drop at the film solution interface.

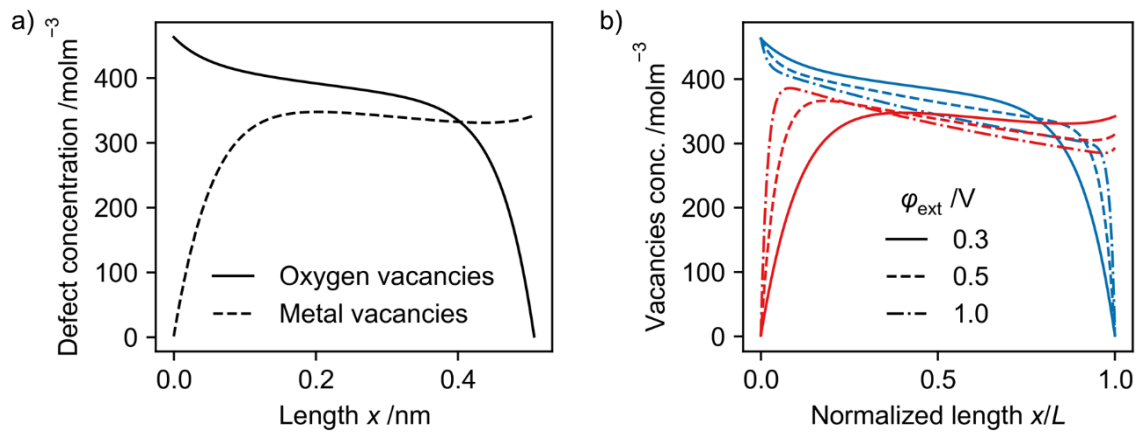

Figure 7: Vacancies concentration depending on the external potential; a) Vacancies distribution over the oxide film for an external potential of 0.3 V; b) Vacancies concentration over the normalized film length; blue: oxygen vacancies, red: metal vacancies.

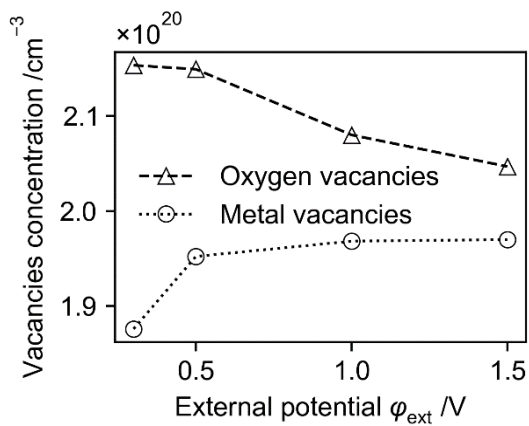

Figure 8: Average vacancies concentration depending on the external potential.

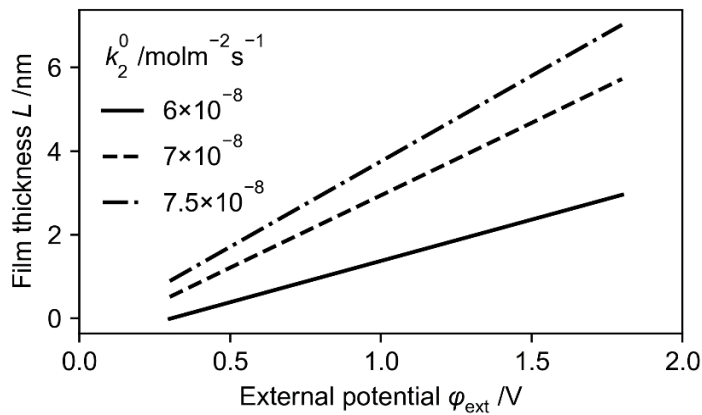

Figure 9: Influence of base rate of oxygen production $k_2^0$ on the steady state oxide film thickness depending on the external potential.

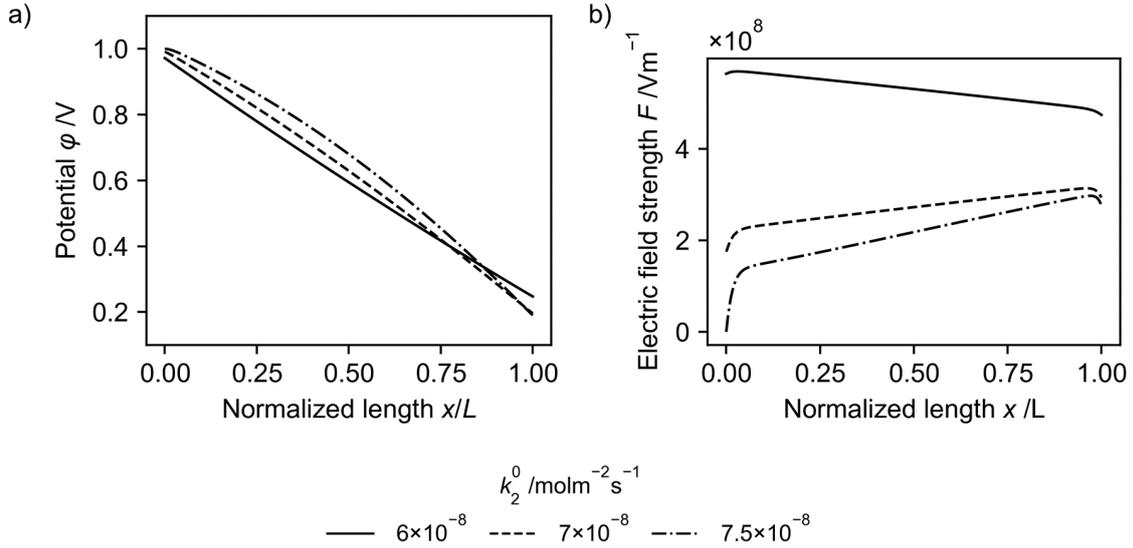

Figure 10: a) Potential distribution over the normalized film length; b) Electric field strength over the normalized film length.

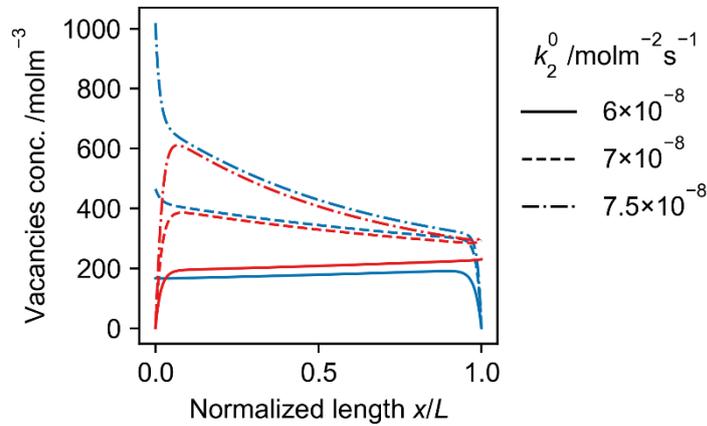

Figure 11: Vacancies concentration over the normalized film length for different standard anion vacancies production rates $k_2^0$; blue: oxygen vacancies, red: metal vacancies.

The higher the base rate of oxygen production $k_2^0$, the more the potential distribution differs from a linear course over the oxide film (Figure 10a). The curvature of the potential distribution is dictated by the electric field strength (Figure 10b). If the number of oxygen vacancies is higher compared to the number of metal vacancies, the electric field strength increases from the metal/film interface toward the film/solution interface. If the number of metal vacancies exceeds the number of oxygen vacancies, the electric field strength is higher at the metal/film interface. This is the case if the steady state rate of metal vacancies production is higher compared to the steady state rate of oxygen vacancies production (assuming similar consumption rates), which is the case for $k_2^0 = 6 \times 10^{-8}$ mol/m²/s[1] (Figure 11). It is important to note that the steady state rates of both reactions are affected by the standard reaction rates and by the potential drops at the corresponding interfaces (mf-interface in case of POV and fs-interface in case of PMV), which are affected among others by the film thickness, the defect concentration, the external potential and the concentration of ions in the solution at the oxide film surface.

## 3.3  Influence of Solution Concentration

The potential drop at the film solution interface drives the electrochemical reaction occurring at both sides of the interface between oxide film and solution (in the studied case there are only electrochemical reactions on the oxide film side of the interface). The potential drop is compounded of the potential drop over the defect layer and over the compact double layer (Figure 3a). This region remains without localized charges and thus, the electric potential at the oxide film/defect layer boundary as well as the potential at the compact double layer/solution boundary dictate the electrochemical reactions.

With increasing ion concentration in the solution, the film thickness of the oxide film increases as well (Figure 12a). This is due to increasing potential drops at the film/solution interface (Figure 12b) which accelerates the electrochemical reactions, increases the defect concentration and thus the electric field. This in turn increases the transport of defects and leads to a steeper decrease of the potential over the oxide film (Figure 13a). Additionally, the higher ion concentration leads to a less pronounced diffuse double layer in the solution and the potential inside the solution is more buffered and reaches the bulk solution value of 0 V faster (Figure 13b).

If the potential drop at the film/solution interface would be known a priori, it might not be necessary to include the solution to the calculation. But since the potential drop at this interface is also a function of the other model parameters, involving the solution to the calculation is recommended. As can be seen, the influence of ion concentration in the solution on the film thickness as well as on $\varphi_{\text{fs}}$ shrinks with higher ion concentration, which is due to a higher potential drop in the compact double layer and less potential drop inside the diffuse layer. With very high ion concentration the assumption of a complete potential drop to the counter electrode potential (in this case 0 V over the defect layer and the compact double layer can be justified. Nevertheless, it is important to implement the electric flux condition to this potential drop.

The potential inside the solution and the concentration distribution of the ions in the solution show the expected behavior due to the Gouy-Chapman-Stern theory (see Appendix).

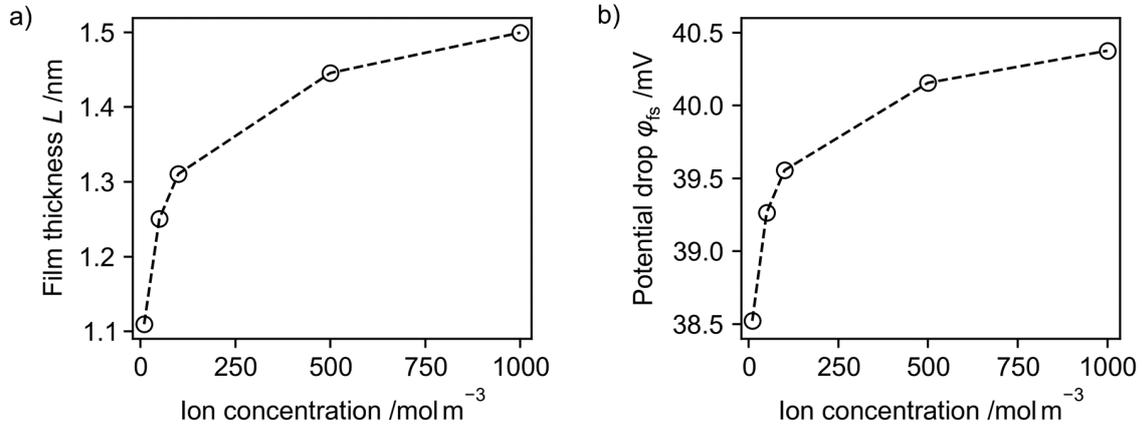

Figure 12: Influence of ion concentration in the solution at the oxide film surface on a) the steady state film thickness and b) the potential drop at the film/solution interface. The potential drop at the film solution interface is calculated from the oxide film/defect layer interface toward the compact double layer/solution interface (compare Figure 3a) and drives the electrochemical reaction at the film/solution interface. The calculations are performed with an external potential of $\varphi_{ext}$ 0.5 V.

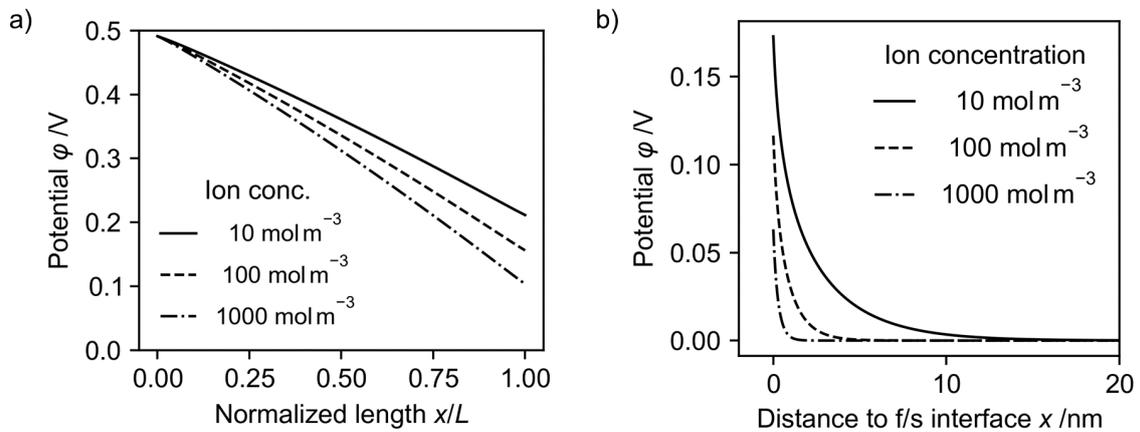

Figure 13: The calculations are performed with an external potential of $\varphi_{ext}$ 0.5 V.

## 4. COMPARISON TO THE CLASSIC PDM

The inclusion of the Poisson equation to calculate the electric field strength for the modeling of passive films leads to significant differences compared to the classic PDM. The steady state film thickness $L_{ss}$ based on the classic PDM can be calculated by [33]:

$$L_{ss} = \frac{1-\alpha}{F_E}\varphi_{ext} - 1/(2\alpha_2 F_E \gamma)\ln\left(\frac{k_5}{k_2^0}\right) \quad (14)$$

assuming a constant pH (the pH dependent term can be included to $k_2^0$), $\gamma = F/R/T$ and $\alpha$ is the polarizability of the film/solution interface. By $\alpha$ the potential drop at the film/solution interface becomes potential dependent (by the R-PDM calculation this potential drop is calculated from the model and no parameter needs to be assumed). It should be noted that $L_{ss}$ calculated by the original PDM is independent of the reaction constants of the oxygen vacancies consumption rate and the metal vacancies production and consumption. In case of the R-PDM the concentration of the metal and oxygen vacancies affects the electric field and thus $L_{ss}$.

The steady state film thickness calculated by the original PDM and the R-PDM is shown in Fig. 14. The chosen values are identical to the R-PDM parameters (if possible), additional parameters are listed in Table 2. The electric field strength $F_E$ has been chosen to be similar to the calculated field strength by the R-PDM (even though it is an estimated value in the PDM). As can be seen the course of the steady state film thickness strongly depends on the voltage dependency of the potential drop at film solution interface $\varphi_{fs}$ - the polarizability of the film/solution interface $\alpha$ (Fig. 14a), a behavior which is automatically calculated by the R-PDM. Depending on the choice of $\alpha$ the estimated steady state film thickness calculated by the PDM can strongly differ from the R-PDM (dotted line Fig. 14a) or can show a very similar course (dashed line Fig. 14a).

A calculation of the defect concentrations can also be done by the classic PDM. According to the Nernst-Planck equation the flux of defects can be described by:

$$J = -D\frac{\partial c}{\partial x} + \frac{zFD}{RT}F_E c \quad (15)$$

Assuming a nearly constant defect concentration [10] (a simplification which is not generally valid according to the R-PDM) one can estimate the oxygen vacancies concentration at the metal film boundary by:

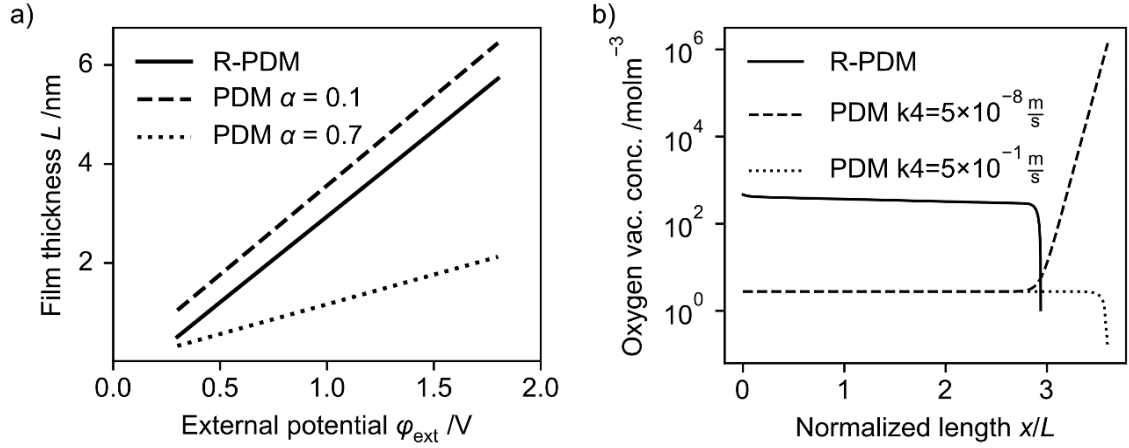

Figure 14: Comparsion of classic PDM calculations to R-PDM calculations; a) Steady state film thickness, with two different polarizabilities of the film/solution interface; b) Oxygen vacancies concentration, with two different vacancies consumption rates in case of the classic PDM.

$$J_{ss} = k_2 = \frac{2FD_{AV}}{RT}F_E c_{AV}^{mf}, \tag{16}$$

which leads to

$$c_{OV}^{mf} = \frac{k_2 RT}{2FDF_E}. \tag{17}$$

The concentration on the film solution interface $c_{OV}^{fs}$ is given by the steady state condition:

$$k_2 = c_{OV}^{fs} k_4. \tag{18}$$

The solution of the differential equation leads to the concentration profiles of the oxygen vacancies across the passive film (Fig. 14b)), which in contrast to the steady state film thickness $L_{ss}$, strongly depend on the rate of oxygen vacancies consumption $k_4$. A similar rate of oxygen vacancies consumptions compared to the production rate (dashed line) leads to a clear increase of $c_{OV}$ towards the film/solution interface It should be mentioned that the concentration profile depends on both reaction rates and by a choice of a significant higher standard reaction rate $k_2^0$ – around 6 orders of magnitude – the concentration drops toward the film/solution interface even though the reaction rates $k_2$ and $k_4$ (given by the standard reaction rates) are similar. If the consumption rate is 7 orders of magnitude higher, the concentration profile corresponds to the expected course, $c_{OV}$ drops towards the film solution interface (dotted line). Other PDM publications (e.g. [27]) also show these large differences between reaction rates leading to the concentration profile. An uneven concentration profile at the metal/film interface, as calculated by the R-PDM (solid line) cannot be achieved by the classic PDM (on the presented way). Furthermore, the PDM calculation based on the chosen parameters lead to significantly lower defect concentrations inside the film.

Table 2: Model parameter for classic PDM calculations (if different from R-PDM parameters)

| Symbol | Model parameter | Value | Unit |
|---|---|---|---|
| $k_4^0$ | Base rate constant of Reaction 4 (COV) | $5 \times 10^{-8} / 5 \times 10^{-1}$ | m/s |
| $F_E$ | Electric field strength | $2.5 \times 10^8$ | V/m |
| $\alpha$ | Polarizability of film/solution interface | 0.1/0.7 | |

## 5. Conclusion

A comprehensive model for oxide growth on metal electrodes is shown. The model replaces debatable assumptions by equations for charge carrier transport including physically valid boundary conditions and by consideration of the impact of the chemical environment.

For the solution of the Poisson equation new boundary conditions are introduced. They are based on proposing two, non-charged defect layers (one at the metal-film and one at the film-solution interface) and on involving the compact double layer; all of finite thickness. These boundary conditions allow a detailed description of (1) the potential drops at the metal-film and film-solution interface, (2) the potential profile in the film and (3) the potential drop in the compact double layer. The thickness of these layers are additional parameters (depending on the lattice constants and the electrolyte concentration) but replace other PDM parameters ($\alpha, \beta, \varphi_{fs}^0$). As a result, several potential drops can be observed at the different interfaces, which drive the electrochemical reactions that occur at these interfaces. The model includes the transport and reactions of defects inside the oxide film, the calculation of potential and electric field inside the film by the Poisson equation, as well as the potential and ion distribution inside the solution at the oxide film surface.

Several conclusions can be drawn from the comprehensive model:
- The assumption of a potential independent electric field strength is not generally valid.
- The Refined-PDM, which is not limited by such an assumption, is more broadly applicable and free of physically unsound boundary conditions.
- The assumption of a constant field strength is only valid in very specific cases.
- The potential drop at the film/solution interface $\varphi_{fs}$ shows a dependency on the external potential.
- It is necessary to include the environment (namely the potential drop across the compact double layer) to the model for the correct calculation.
- The choice of model parameters strongly affects the outcome of the calculation

and several different oxide films can be shown by the model.

As already discussed in the text, the PDM hypothesis of potential independent electric field strength is based on the assumption of tunneling current and there can be rare cases at which this affects the electric field strength. A detailed description of this cases will be explored in the future.